\documentstyle[epsf]{mn}

\title[Hybrid $C_{\ell}$]
{The hybrid SZ power spectrum: Combining cluster counts and SZ fluctuations to probe gas physics}

\author[J.M. Diego \& S. Majumdar]
   {J.M. Diego$^1$ \& Subhabrata Majumdar$^2$\\
    $^1$University of Pennsylvania.
    David Rittenhouse Laboratory, 209 South 33rd St,\\ 
    Department of Physics \& Astronomy, Philadelphia PA 19104, US.\\
    $^2$ CITA, University of Toronto, 60 St George st, Ontario, M5S3H8,    
    Canada\\
    } 
\date{Draft version \today}

\pagerange{\pageref{firstpage}--\pageref{lastpage}}

\begin{document}

\maketitle

\label{firstpage}
\begin{abstract} 

Sunyaev-Zeldovich (SZ) effect from a cosmological distribution of clusters
carry  information on the underlying cosmology as well as the cluster gas
physics. In order to study either cosmology or clusters one needs to break the
degeneracies  between the two. We present a toy model showing how complementary
informations from SZ power spectrum  and the SZ flux counts, both obtained from
upcoming SZ cluster surveys,  can be used to mitigate the strong cosmological
influence (especially that of $\sigma_8$) on the SZ fluctuations.  Once the
strong dependence of the cluster SZ power spectrum on $\sigma_8$ is diluted, 
the cluster power spectrum can be used as a tool in studying cluster
gas  structure and evolution. The method relies on the ability to write the
Poisson contribution to the SZ power spectrum in terms  the observed SZ flux
counts.  We test the toy model by applying the idea to simulations of SZ
surveys. 

\end{abstract}

\section{Introduction}\label{section_introduction}

In the last decade, X-Ray observations of galaxy clusters have been used 
extensively to determine  the cosmological matter density parameter and the
amplitude of density fluctuations (Henry 1997, Bahcall \& Fan 1998,  Viana \&
Liddle 1999). In near future, advent of new mm and sub-mm high-sensitivity
experiments will open  a new window for studies of galaxy clusters through the
SZ    effect surveys.  These surveys  (for example
SPT\footnote{http://astro.uchicago.edu/spt/}, 
ACT\footnote{http://www.hep.upenn.edu/$\sim$angelica/act/act.html},
APEX-SZ\footnote{http://bolo.berkeley.edu/apexsz/}) would be capable,  through
wide area coverage and high sensitivity, of detecting  thousands of galaxy
clusters up to  redshift $z > 1$. The cosmological possibilities of such a
large data sets are  enormous and will allow to carry out independent
estimations of the  cosmological parameters (Diego et al. 2002, Levine et al.
2002, Weller 2002, Majumdar  \& Mohr 2003a,2003b, Hu \& Kravtsov 2003, Lima \&
Hu 2004) which could be compared with those  obtained from  other observations
(CMB, SNIa, Ly-$\alpha$ forest, etc).  Specifically, the cluster abundance with
redshift, $dN\over  dz$,  would provide constraints on the dark energy
equation  of state parameter $w\equiv p/\rho$ (Wang \& Steinhardt 1998).  Since
SZ effect is redshift independent, to get the redshifts of the clusters 
detected in these SZ surveys, one would need to have followup observations in
the optical/IR bands. In the absence of redshifts, a large yield SZ
survey have cluster number counts as a function  of SZ flux, $\mathcal{N}(S)$ would be available.
This is also sensitive to cosmological parameters (Barbosa et al. 1996, 1998,
Bartlett 2000) but weaker in comparison to the redshift abundance. It is
interesting then to explore alternative methods which exploit  the new SZ data
sets without any need of $z$ information. This is the main essence of the
paper. 

In the absence of any redshift information one would also be able to
estimate  the SZ power spectrum from the temperature maps in addition to the
flux counts. The statistics of SZ maps have also been well studied, both
analytically (e.g., Cooray 2000, Zhang \& Pen 2001, Komatsu \& Seljak 2002) and
through  numerical simulations (e.g., Refregier et al. 2000, Seljak et al. 2001,
Springel et al. 2001, da Silva et al. 2001, Zhang, Pen \& Wang 2002). Comparison
made between the two approaches (Refregier \& Tessier 2002, but also see Zhang
et al. 2004)  show that a simple halo model (Cooray \& Sheth 2002) description
of the clusters reproduces results from detailed N-body simulations. Some
differences remain, especially at very high $\ell$'s where presumably the 
effect from substructures creep in. All these studies come to  similar
conclusions about the potential of using the cluster SZ power spectrum to
constrain the background cosmology, especially $\sigma_8$ (Bond et al. 2002). Apart from cosmology, it
has also been shown that with some knowledge of cosmological parameters, the
SZ power spectrum can be used as a powerful probe of cluster gas physics
(Majumdar 2001, Holder \& Carlstrom 2001, Sadeh \& Rephaeli 2004a).  It is
clear from all these studies that the possibilities of doing science with the
SZ power spectrum are enormous.

It is, however, well known that cosmological studies based on cluster  data
will show degeneracies in the cosmological parameters and  cluster scaling
relations. One of the best known is the degeneracy between  $\sigma _8 - \Omega
_m$ which can only be weakened if one can measure the  evolution of the cluster
population as a function of redshift.  Moreover, imperfect knowledge of cluster
structure and evolution can weaken cosmological constraints obtained from these
surveys. In the presence of any such evolution of gas physics mimicking
cosmological evolution of the cluster number density, one would have to resort
to other options  (Majumdar \& Mohr 2003b, Lima \& Hu 2004) to tighten the constraints. In the
absence of  any redshift information, when dealing with flux counts or power
spectrum, one encounters similar degeneracies  between cosmology and cluster
physics. One can try to break these degeneracies by directly adding constraints
from flux counts to those from power spectrum 
resorting to time intensive joint-fits to the two data sets. This has the
advantage in that it not only gives us the constraints on the different
parameters but also the correlations between them. However, it is instructive
to  look whether at all one can disentangle cosmology from cluster physics
before resorting to more time consuming techniques. It is in this `exploratory'
sense that we try to look at the possibility of mitigating the cosmological
influence by combining flux counts to SZ power spectrum such that it becomes
possible to probe gas physics. The intention is not to estimate
precise constraints achievable from upcoming SZ surveys but to look
at the feasibility of studying cluster gas physics with these surveys. This approach towards studying cluster properties would be worthwhile since the present uncertainties in determining cluster properties from targetted observations remains large (like upto many tens of percents for any redshift evolution of the cluster scaling relations).

In this work, we propose to combine the cluster number counts as a function  of
SZ effect flux, $\mathcal{N}(S)$, and the power spectrum of the sample,
C$_{\ell}$,  in such a way such as to dilute the dependency of the SZ power
spectrum on cosmological parameters.  Each depends on the background cosmology
and gas physics in slightly different way than the other (for example, the SZ
power spectrum depends on the detailed distribution of the gas in a cluster
whereas the SZ flux only depends on the total thermal content of the cluster). 
Although both $\mathcal{N}(S)$ and $C_\ell$ depend on the total fluxes of the
clusters, the latter also depends on the shape of the clusters, the size of
cluster being inversely proportional to its angular distance. Thus the
cosmology-gas physics degeneracies are different for the two quantities. 
Let us note, once again, at this point that
both  $\mathcal{N}(S)$ and $C_\ell$ are quantities where the implicit redshift
dependence is integrated over unlike that from $\frac{dN}{dz}$. Moreover, single
SZ survey would give us both the quantities and hence no extrapolation is needed
between observations at different wavelengths.

The rest of the paper is organized as follows: in \S \ref{sec_formalism} we lay down 
the basic formalism and then in \S  \ref{sec_modelling} we describe the toy model of the 
clusters. The standard SZ power spectrum and its dependences on different parameters 
are shown in \S \ref{sec_powspec}. We present the simulations in \S \ref{sec_simulations} 
and in \S \ref{sec_clusterphys} we show the use of hybrid power spectrum as a tool 
for probing gas physics and discuss the systematics. 
Finally, in  \S \ref{sec_summary} we discuss and summarize the method and results.

\section{The hybrid cluster power spectrum: Formalism}
\label{sec_formalism}

The total SZ power spectrum from a redshift distribution of clusters, biased
w.r.t. the underlying dark matter distribution, can be written as the sum of
two terms, the 1-halo Poisson term and the 2-halo clustering term. 
In general, the clustering term is dominant for $\ell < 100$ (Komatsu \&
Kitayama 1999) where the shape of the power spectrum is determined mainly by the 
distribution of the halos. This region is  driven mainly by the
cosmological model. On the other hand, the  shape of the individual gas
profiles determine the higher $\ell$-values (around and beyond the peak of the
total SZ power spectrum). Thus, this part is more influenced by cluster
physics. Taken together, there remains a large amount of degeneracy between
the gas physics and cosmology.   

The Poissonian contribution to SZ power spectrum, $C^P_l$, from galaxy clusters can be 
described as,
\begin{equation}
C^P_l = \int dz \frac{dV(z)}{dz} \int dM \frac{dn(M,z)}{dM} p_l(M,z) ,
\label{eqn:Clp}
\end{equation}
where $dV(z)/dz$ is the comoving volume element, $dn(M,z)/dM$ is the cluster mass 
function (Press \& Schechter 1974, Sheth \& Tormen 1999, Jenkins et al. 2001) 
and $p_l(M,z)$ is the power spectrum (multipole decomposition) of the 
SZ effect 2D profile of a cluster with mass $M$ at redshift $z$.

Contribution to the total SZ power spectrum  arising from the spatial correlations of
the clusters, $C^C_l$ is given by
\begin{eqnarray} 
C^C_l = &&\int dz {{dV(z)}\over{dz}} P_m \times  \nonumber \\
&& {\left[\int dM {{dn(M,z)}\over{dM}}  b(M,z)
p_l^{1/2}(M,z)\right]} ^2 ,
\label{eqn:Clc} 
\end{eqnarray} 
where $b(M,z)$ is the time dependent linear
bias factor.  The matter power  spectrum, $P_m(k,z)$, is related to the power
spectrum of cluster correlation function  $P_c(k,z)$ through the bias, i.e.,
$P_c(k,z)=b^2(M,z)D^2(z)P_m(k,z=0)$ where we  adopt  
$b(M,z) =1 + (\nu^2(z) -1)/\delta_c(z)$ (Mo \& White 1996).  
In the above equation $D(z)$ is the linear
growth factor of density fluctuation, $\delta_c(z=0)=1.68$ and 
$\nu(z)=\delta_c(z) /\sigma(M)$.

As we have already mentioned, future SZ surveys would detect many thousands of
galaxy clusters giving us a flux limited sample  $\mathcal{N}(S)$ of clusters. 
Since cluster detection does not care about the spatial correlations of the 
clusters, hence $\mathcal{N}(S)$ is simply the  Poisson
distributed sample of clusters. It is easy to see that the SZ contribution from
this sample would give us the 1-halo SZ power spectrum once we are careful in
evaluating eq. (\ref{eqn:Clp}). One may then try to rewrite eq. (\ref{eqn:Clp}) 
in terms of $\mathcal{N}(S)$. 

To proceed let us write equation (\ref{eqn:Clp}) 
in the following form,
\begin{equation}
C^P_l = \int dz \int dS \ N(S,z) p_{\ell}(S,z)
\label{eqn_Cl_b}
\end{equation}
where the term $N(S,z)$ is the number of clusters with flux $S$ and at 
redshift $z$ per redshift and flux intervals.
In the previous equation, instead of the masses of the cluster, we have used  
the measured SZ effect flux (see next section). 
The term $p_l(S,z)$ is the power spectrum of a cluster with flux $S$ at 
redshift $z$.
\begin{equation} 
p_{\ell}(S,z) = \frac{S^2}{4 \pi} f_{\ell}(S,z)
\label{eqn_pl}
\end{equation}
The function $f_{\ell}(S,z)$ gives the shape of the power spectrum as 
a function of $\ell$. These terms are model dependent and depends sensitively 
on the cluster structure and its evolution.  \\

Combining equations (\ref{eqn_Cl_b}) and (\ref{eqn_pl}) we get the expression 
for the {\it hybrid} power spectrum, $C^H_l$, as
\begin{equation}
C^H_l = \int dS\  \frac{S^2}{4\pi} \int dz \  N(S,z) f_{\ell}(S,z)
\label{eqn_Cl_h}
\end{equation}

The trick is to express $N(S,z)$ as a function of the observed 
$\mathcal{N}(S)$ (observed number of clusters with fluxes in the interval 
$[S - \Delta S/2, S + \Delta S/2]$). This is possible if we use our  understanding 
of the mass function, $dn/dM$, of dark matter halos from theory and simulations. 
One can use scaling relations between the mass and SZ flux to rewrite the mass 
function in terms of the observed SZ fluxed.

Thus, we have 
\label{eqn_NS}
\begin{equation}
N(S,z) = {\mathcal{N}} {\rm(S)} \frac{\Delta z \frac{dV}{dz}\int _{\Delta S}
\ dS \frac{dn}{dS} }{n_{th}(S)}
\label{eqn_nSz}
\end{equation}
where $n_{th}(S)$ is the total (theoretical) number of clusters with fluxes in the interval 
$[S - \Delta S/2, S + \Delta S/2]$. 
\begin{equation}
n_{th}(S) = \int dz \frac{dV}{dz}\int _{\Delta S} \ dS \frac{dn}{dS}
\label{eqn_nTS}
\end{equation}

Equations (\ref{eqn_Cl_h}) and (\ref{eqn_nSz}) condense the main philosophy of 
our method. Build the SZ power spectrum as a function of the flux 
function instead of the mass function (equation \ref{eqn_Cl_h}) and use 
a normalized version of the flux function (equation \ref{eqn_nSz}) such 
that its integral in redshift is the observed number counts, that is:
\begin{equation}
{\mathcal{N}} {\rm(S)} = \int dz N(S,z) 
\end{equation}

Equation (\ref{eqn_nSz}) is particularly interesting because it contains the 
cosmological dependence in both the volume element $dV/dz$ and the number 
counts $dn/dS$ which is connected to the mass function through the chain rule.
\begin{equation}
\frac{dn}{dS} = \frac{dn}{dM} \frac{dM}{dS}
\label{eq_chain}
\end{equation}
where $dn/dM$ is the cosmology-dependent mass function. The above expression
also  shows how $dn/dS$ depends on the $S-M$ relation. We will see in the next
section how  this relation depends on the $T-M$ scaling relation. Summing up,
the cosmological  dependency of the hybrid power spectrum will appear only in
the $N(S,z)$ term but it will  be diluted because of the normalization by the
factor $n_{th}(S)$ in equation (\ref{eqn_nSz}).  On the other hand, the
dependency on the gas physics (or equivalently in the scaling  relations) will
appear in $N(S,z)$ through equation (\ref{eq_chain}), in the  $S^2$ term of
equation (\ref{eqn_Cl_h}) (because we need to transform masses into fluxes)  and
in the power spectrum of a cluster with flux $S$ at redshift $z$, $f_l(S,z)$
as  we will see in the next section. It turns out that the gas physics 
dependence is not diluted as much as the cosmological one.

To model the number density of clusters as a function of redshift and mass,  we
use the standard Press-Schechter mass function although one is free to choose
other mass functions  which render better fits to numerical N-body simulations.
The point to be noted is that in writing the hybrid  SZ power spectrum we have
essentially normalized the cluster Poisson power spectrum to the observed flux
limited sample of clusters. As we will see, this has interesting implication
for the use of cluster SZ $C_l$ as probes of cosmology and cluster physics. The
resultant power spectrum has weaker dependence on cosmology (for example,
$\sigma_8$) while still retaining the sensitivity to gas physics.



\section{Modeling the galaxy cluster }
\label{sec_modelling}

To describe the cluster structure and evolution, we use a simple model which can
fit current observations (Diego et al. 2001). 
To describe the SZ profile of a cluster one requires only three basic ingredients:  
the temperature of the clusters, their sizes and geometry (i.e radial profile). 
Based on X-Ray observations (Verde et al. 2001, Finoguenov 2001), the  
temperature and size are modeled using the scaling relations 
\begin{eqnarray}
T &=& T_o M^{\alpha}(1 + z)^{\phi}  
\label{eqn_TM}\\
R_v &=& R_o M^{\beta}(1 + z)^{\psi} ,
\label{eqn_RM}
\end{eqnarray}
where $R_v$ and $T$ are the virial radius and the isothermal temperature of the 
cluster.
In the above relations we have not only defined cluster structure but have also 
assumed a simple redshift evolution for the cluster structure. 
The basic structure and evolution parameters are taken from Diego et al..(2001).

\begin{figure}
   \begin{flushleft}
   \epsfysize=6.cm 
   \begin{minipage}{\epsfysize}\epsffile{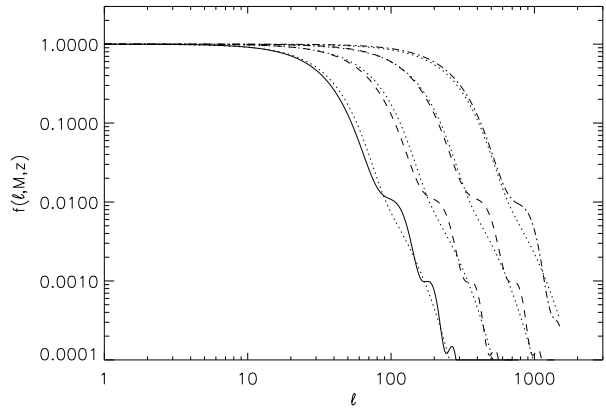}\end{minipage}
   \caption{
            The solid lines give the (normalized) power spectrum ($f_{\ell}(M,z)$) 
            for various clusters models with (from left to right) 
            $r_c = 25, 12, 6$ and 3 arcmin respectively. 
            Dotted lines are the best fitting model 
            given by equations (\ref{eqn_fit1}) and (\ref{eqn_fit2}).
            The bumps in the tails are due to the truncated nature of the 2D profile 
            (we integrate the SZ distortion in a sphere or $R < R_v$).
           }
   \label{fig_flMz}
   \end{flushleft}
\end{figure}

The third ingredient of the model, the geometry, gives the shape of the 
3-D distribution of the electrons in the intracluster medium. 
For this we assume a standard $\beta$-model with $\beta = 2/3$ to describe 
the spatial distribution of the electron density, $n_e(R)$, such that
\begin{equation}
n_e(R) = \frac{n_e(0)}{1+(R/R_c)^2} .
\end{equation} 
This particular model has two free parameters: the core radius, $R_c$, which we take 
from equation (\ref{eqn_RM}) assuming that the ratio between the virial 
and core radius is a constant ($R_v/R_c = 10$). The second free parameter is 
the  normalization. Instead of normalizing with the central electron density, 
$n_e(0)$, we normalize to the total SZ signal (which is an observable) from 
the cluster given by 

\begin{equation}
\Delta T_{tot} (\mu K) = \Delta T_o \frac{T M}{D_a^2} ,
\label{eqn_totalT}
\end{equation}
where the constant $\Delta T_o$ contains the baryon fraction, $f_b$, 
today's mean CMB temperature, $T_{CMB}$ and the frequency dependency of the 
SZ thermal effect, $g(x)$, such that 
\begin{equation}
\Delta T_o = 162 g(x) f_b T_{CMB} \ \ \mu {\rm K}  \ {\rm strad^{-1}}
\end{equation}

In the above equations, the temperature is given in Kev, the 
mass in $10^{15} h^{-1} M_{\odot}$, the baryon fraction in $h^{-1}$ and the 
angular diameter distance, $D_a$, in $h^{-1}$ Mpc. 
The $h$-dependency of the mass and baryon fraction cancel outs with the one 
in $D_a$. The frequency dependence, $g(x)$ is given by

\begin{equation}
g(x) = \frac{x}{{\rm tanh}(x/2)}-4 ,
\end{equation}
where $x =  \nu({\rm GHz})/56.8$.

Although, SZ power spectrum is quoted in $\mu {\rm K}^2$, the cluster fluxes are
quoted in mJy. Therefore, the corresponding expression for the total flux would be

\begin{equation}
S_T({\rm mJy}) = S_o \frac{T M}{D_a^2}.
\label{eqn_totalS}
\end{equation}
As above, $S_o$ contains all the relevant constants and is given by
\begin{equation}
S_o = 4.4 \times 10^7 f_x f_b \ \  {\rm mJy}
\end{equation}
where $f(x) = g(x)x^4 e^x /(e^x -1)^2$ .

To calculate the cluster power spectrum (see equation \ref{eqn_Cl_b}) 
we also need an expression for the form factor $f_{\ell}(S,z)$  
appearing in equation (\ref{eqn_pl}). This is obtained by fitting 
the  Fourier transforms of SZ profiles of clusters from simulations where the mass 
and redshift of the cluster are changed (keeping $R_v/R_c$ constant).
A fit of this kind using our model to simulate the clusters gives;
\begin{equation}
f_{\ell}(M,z) = 0.9[{\rm exp}(-\xi_{l,r_c})] + 0.1[{\rm exp}(-\sqrt{\xi_{l,r_c}})]
\label{eqn_fit1}
\end{equation}
with, 
\begin{equation}
\xi_{l,r_c} = l^2 r_c^{1.5/(0.92 + 10.4r_c)}
\label{eqn_fit2}
\end{equation}
where the core radius, $r_c$, is given in radians. 
Note that in the limit of very large angular scales (i.e $\ell \rightarrow 0$) we 
have $f_{\ell}(M,z) \rightarrow 1$ and the cluster behaves as a point source
having a constant power spectrum. 
In figure (\ref{fig_flMz}) we show the quality of our fit for various cluster models 
where we have changed the core radius.

\section{Cosmological and physical dependency of the standard power spectrum.}
\label{sec_powspec}

In order to compute the hybrid power spectrum we need the 
observed curve ${\mathcal{N}}(S)$. 
The flux limits of the source counts are determined by the sensitivity limit 
of the survey giving $S_{min}$. The maximum flux $S_{max}$ is determined by the flux
of the brightest cluster in the catalog. 
The curve ${\mathcal{N}}(S)$ is usually affected by the selection 
function of the survey, that is, below a certain flux, not all the 
clusters are detected.  
In our case we will consider that all these points have been included 
and that the final ${\mathcal{N}}(S)$ is an unbiased estimate 
of the underlying cluster number counts. To simulate the cluster distribution, 
we assume that the number of clusters in the 
flux bin $i$ (that is, clusters with fluxes $_i \in [S_i,S_{i+1}]$) is a 
Poissonian variable with mean value $\bar{S}_i$ where $\bar{S}_i$ 
is the expected number of clusters in that bin for a given model. 
The error bars assigned to each bin are Poissonian error bars. With flux counts determined, we 
are ready to compute the different power spectra. To understand the dependencies 
of ${\mathcal{N}}(S)$ and $C^P_l / C^H_l$ on the underlying cosmological models 
and gas physics, we look at the following observationally motivated models, given in Table 1.

\begin{table} 
 \label{table1}
 \begin{center}
  \begin{tabular}{|c|c|c|c|c|c|c|c|c|c|}
    \hline
    \hline
     &$\Omega _m$&$\sigma _8$&$\Lambda$&$T_o$&$\alpha$&$\phi$&$R_o$&$\beta$&$\psi$ \\
    \hline
    \hline
     A&0.3&0.8&0.7&8.0&2/3&1.0&1.3&1/3&-1\\
    \hline 
     B&{\bf 0.5}&0.8&0.7&8.0&2/3&1.0&1.3&1/3&-1\\
    \hline 
     C&0.3&{\bf 1.0}&0.7&8.0&2/3&1.0&1.3&1/3&-1\\
    \hline 
     D&0.3&0.8&{\bf 0.0}&8.0&2/3&1.0&1.3&1/3&-1\\
    \hline 
     E&0.3&0.8&0.7&{\bf 10.0}&2/3&1.0&1.3&1/3&-1\\
    \hline 
     F&0.3&0.8&0.7&8.0&{\bf 1/2}&1.0&1.3&1/3&-1\\
    \hline 
     G&0.3&0.8&0.7&8.0&2/3&{\bf 0.0}&1.3&1/3&-1\\
    \hline 
     H&0.3&0.8&0.7&8.0&2/3&1.0&{\bf 1.0}&1/3&-1\\
    \hline 
     I&0.3&0.8&0.7&8.0&2/3&1.0&1.3&{\bf 1/4}&-1\\
    \hline 
     J&0.3&0.8&0.7&8.0&2/3&1.0&1.3&1/3&{\bf 0}\\
    \hline 
  \end{tabular}
  \label{table:models}
  \caption{Models used to illustrate the sensitivity of the hybrid power 
           spectrum to the cosmological model and the cluster physics. 
           Model A is the fiducial model. The numbers in bold face  
           show the parameter which has been changed compared to the fiducial   
	   model. All numbers are dimensionless except $T_o$ (Kev) and $R_o$  
	   ($h^{-1}$ Mpc)}
 \end{center}
\end{table}

In the table, model A is our fiducial model. We have nine free parameters of 
which three relates to the underlying cosmology (whose changes w.r.t the 
fiducial model are shown in models B,C \& D) and the other six parameters 
define cluster structure and evolution (changes to these parameters shown 
in model E,F,G,H,I \& J).

\begin{figure}
   \begin{flushleft}
   \epsfysize=6.cm 
   \begin{minipage}{\epsfysize}\epsffile{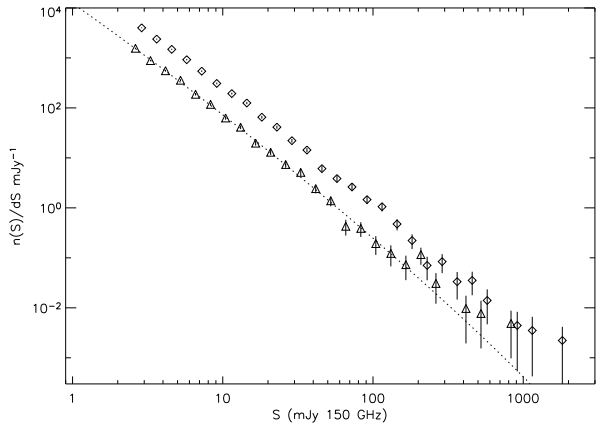}\end{minipage}
   \caption{
            Simulated ${\mathcal{N}}(S)$ curves 
	    for an experiment with the 
            characteristics of ACT (150 sq. deg area, and a few mJy 
            minimum flux at 150 GHz)
	    for two models with 
            $\sigma_8 = 1.0$ (top, model C) and $\sigma_8 = 0.8$ 
	    (bottom, model A). The 
            other parameters are like in model A in Table 1 (except for 
            $T_o$ for which we use $T_o = 8.5$ keV). 
             The fluctuations at high 
            fluxes follow Poissonian statistics as well as the error bars. 
	    The dotted line is a fit to model A.
           }
   \label{fig_Ns}
   \end{flushleft}
\end{figure}

In figure \ref{fig_Ns} we show two examples of simulated cluster number  counts
for two models (A \& C) where we only change $\sigma _8$.  We have also fitted 
the  ${\mathcal{N}}(S)$ from model A (shown by smooth dotted line). 
One can use the fit to extrapolate
the model to lower flux limit below the detection sensitivity of the surveys.
A point to be noted is that the two number counts are easily differentiable 
(for example, they are $\sim 8\sigma$ apart at $\nu \sim 25$mJy and $\sim 2\sigma$ 
at $\nu \sim 250$mJy). Since, the flux counts for different cosmologies are easily 
differentiable, it makes sense to normalize the $C_\ell$ using 
${\mathcal{N}}(S)$. In section \ref{sec_clusterphys}, we show how this helps to 
use $C_\ell$ to study gas physics.

\begin{figure}
   \begin{flushleft}
   \epsfysize=6.cm 
   \begin{minipage}{\epsfysize}\epsffile{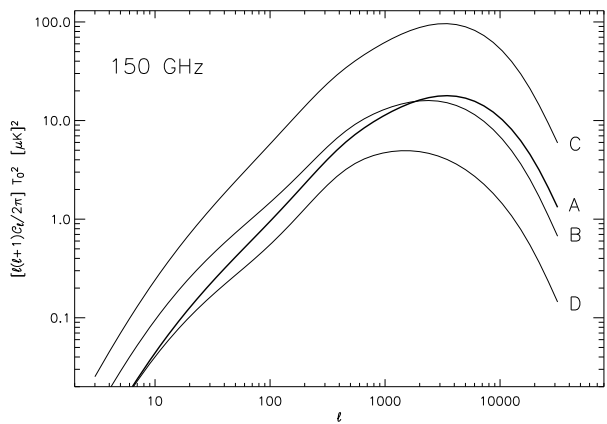}\end{minipage}
   \caption{
            This plot shows the dependence of the standard 
            power spectrum given by equations (\ref{eqn:Clp}) \& 
            (\ref{eqn:Clc}) on cosmological 
            parameters. The models are described in Table 1 and in the text.
           }
   \label{fig_totalcosmo}
   \end{flushleft}
\end{figure}

\begin{figure}
   \begin{flushleft}
   \epsfysize=6.cm 
   \begin{minipage}{\epsfysize}\epsffile{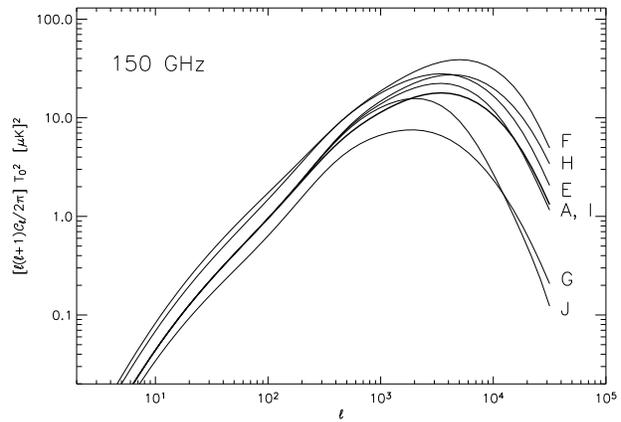}\end{minipage}
   \caption{
            This plot shows the dependence of the standard 
            power spectrum given by equations (\ref{eqn:Clp}) \& (\ref{eqn:Clc})
            on cluster structure and evolution. 
            The models are described in Table 1 and in the text.
           }
   \label{fig_totalphys}
   \end{flushleft}
\end{figure}

Other than the flux counts coming out of the surveys, one can use the
temperature  fluctuations in the survey area to  construct the total power
spectrum, $C^T_l$, (given by the sum of $C^P_l$  and $C^C_l$)\footnote{To avoid
confusion, note that $C^T$ is not the same as $C^H$  which we have introduced
in the text}.  The total power spectrum of SZ effect from galaxy clusters can
be estimated with  considerable accuracy using multifrequency observations of
the CMB  (Hobson et al. 1998, Bouchet \& Gispert 1999, Cooray et al. 2000, 
Maino et al. 2002, Mart\'\i nez-Gonz\'alez  et al. 2003, Delabrouille et al.
2003). 

In figure (\ref{fig_totalcosmo}) we show $C^T_l$ for four models  (A, B, C, D)
which differ in the values of the cosmological parameters. As shown numerous
times in previous studies, the SZ curves are very sensitive to cosmology, 
especially to $\sigma_8$ (compare curves A, B \& C). In fact, the strong
dependence of $C^T$ on $\sigma_8$ has been used to argue for high value of
$\sigma_8 \approx 1 $ based on the CBI observations of temperature anisotropies
at high multipoles (Bond et al. 2002). This value if $\sigma_8$ is high
compared to those obtained from primary CMB data, cluster abundance or weak
lensing studies. The approach of constraining $\sigma_8$ from an estimation of
$C^T$, however, has a caveat.  It assumes that we are sure of the cluster
structure and its evolution.  As we see in next figure, cluster physics can
substantially influence the SZ power spectrum  over large  range in
$\ell$-values.

In figure (\ref{fig_totalphys}) we look at the influence of cluster gas
physics  on the total SZ power spectrum. In comparison to figure
(\ref{fig_totalcosmo}), it is  easy to see that changes in cluster physics can
be a source of substantial systematics  over the entire $\ell$-range of the SZ
power spectrum.  Whereas at low $\ell$, say $\ell \approx 1000$, changes in gas
physics modifies the SZ power spectrum slightly less than that due change in
cosmology, at higher multipoles gas physics has an even stronger effect on
$C^T$. Thus, the SZ power spectrum is quite degenerate w.r.t changes in
cosmology and cluster physics. This degeneracy is strongest between cluster
structural evolution and cosmology. It is easy to understand the strong
degeneracy especially at higher  $\ell$-values: changing $\sigma_8$ strongly
affects the cluster population and thus  changes the amplitude of the power
spectrum whereas changing cluster  physics,  especially the evolution, affects
the high redshift population strongly and affects  the small angular scales.   

Thus in order to probe either cosmology or gas physics from SZ observations
one  would need to  consider the degeneracy between the two.  As an example, a
careful consideration of the mass-temperature relation (i.e gas physics) can
yield a weaker dependence of the SZ power spectrum on $\sigma_8$ (i.e $C^T \sim
\sigma_8^4$) instead of the usual  $C^T \sim \sigma_8^7$ (Sadeh \& Rephaeli
2004b).  The focus of this paper is to look at the feasibility of studying
cluster physics with SZ surveys. To do so one would first need to mitigate the
influence of $\sigma_8$ on the SZ power spectrum. As we will show this can be
done by using the hybrid power spectrum. The first thing to realize is that in
figure (\ref{fig_totalcosmo}), curves A \& C (having  two different $\sigma_8$
corresponds to the two ${\mathcal{N}}(S)$ curves shown in  figure
(\ref{fig_Ns}) whereas, in reality, one would observe only one data set giving
one specific ${\mathcal{N}}(S)$. We will use this fact to dilute the strong
influence of $\sigma_8$ on the hybrid power spectrum.  However, before we
proceed further, we must show that the hybrid power spectrum gives a good
description of the standard SZ power spectrum. In the next section we address
this issue.

\section{Cluster survey simulations and the hybrid power spectrum}
\label{sec_simulations} 

The central idea of the present approach lies in replacing the standard power
spectrum with the hybrid power spectrum.  In the halo model, the total  power
spectrum can be written as the  sum of the Poisson contribution and that from
the spatial correlation of the clusters.  Other than the largest angular
scales, the Poisson power spectrum  dominates heavily over the clustering term
(Komatsu \& Kitayama 1999). Thus above $\ell \sim 200$ the  effective SZ power
spectrum comes totally out of contributions from the Poisson distributed
objects. This total SZ power spectrum would be made out  of contributions  from
halos distributed over a wide range of mass and redshifts. At smaller angular
scales  the halo structures have considerable influence.  Now, in reality,
other than the halos there can be diffuse SZ contributions from low density
filamentary  structures. In past simulations (see Springel et al. 2001, da
Silva et al. 2001, Refregier \& Teyssier 2002) it has already been shown that
the dominant contribution to thermal SZ power spectrum comes from halos (or
regions of high overdensities) such that the analytical halo model calculations
of SZ power spectrum give reasonable results over a wide multipole range. These
comparisons were made by isolating the halos (or overdensities) in the
simulations and comparing the power spectrum obtained from these objects to the
total SZ power spectrum obtained in the simulations. In actual observations one
would only detect halos through their observed SZ fluxes, unlike in the
simulation where one can identify each halo. Moreover, detecting a SZ source
would depend on how well we are able to separate out the SZ-sky from the
observed CMB-sky.  The idea of the present work is to see how well we can
construct the full thermal SZ power spectrum from the flux counts recovered in
a typical survey. We have already laid down the procedure to do this in the
previous sections and in this section we use results from numerical simulations
to see whether the method works.


\begin{figure}
   \begin{flushleft}
   \epsfysize=8.5cm
   \caption{Simulated $10^{\circ}\times 10^{\circ}$ map of SZ survey with ACT characteristics at 145 GHz.}
   \label{fig_ACT_fullsky}
   \end{flushleft}
\end{figure}


\begin{figure}
   \begin{flushleft}
   \epsfysize=8.5cm
   \caption{Recovered SZ map for constructing $\mathcal{N}(S)$. 
   The clusters detected by \small{SEXTRACTOR}\normalsize\  are  marked with ellipses.}
   \label{fig_ACT_sextractor}
   \end{flushleft}
\end{figure}

For concreteness, we focus on observations from simulated ACT survey which we
take to have a flux limit of $S_{lim} \approx 5 $ mJy (145 GHz). Depending on
this sensitivity, the survey would sample only a part of  the total cluster
population having fluxes $S > S_{lim}$. We will consider an area of the sky of
100 sq. deg. which is the expected surveyed area after 5 months of observation
(1 month of integration time). ACT  is a three band experiment with central
frequencies at $\nu$ = 145, 225 and 265 GHz  with $\Delta \nu = 20,25,30$ GHz
respectively. The effective resolutions of 1.7, 1.1, and 0.93 arcmin
respectively at these frequencies. The telescope will be  equipped with a
Bolometer array (MBAC) with imaging capabilities over an area  of $22'\times
22'$. The planned sensitivity of the MBAC  per detector are 300, 500 and 700
($\mu$K sec$^{1/2}$) which for one month of  integration time and 100 deg$^2$
surveyed area renders sensitivities of a few to around 10 $\mu$K. 


In simulating the SZ observations we include four ingredients: i) a realization
of the primary CMB, ii) full hydrodynamical N-body simulation of the SZ which
includes  the correlation between the clusters (from White 2003), iii) the
expected noise level of  ACT in its three bands (145, 225 and 265 GHz) and
iv)    the effect of the antenna beam. We neglect point sources in the present
work and assume  all the bright point sources to be subtracted out. Note, that
presence of point sources may be  a source of considerable noise for future SZ
surveys (White \& Majumdar 2003, Knox, Holder \& Church 2003).

\begin{figure}
   \begin{flushleft}
   \epsfysize=6.cm \begin{minipage}{\epsfysize}\epsffile{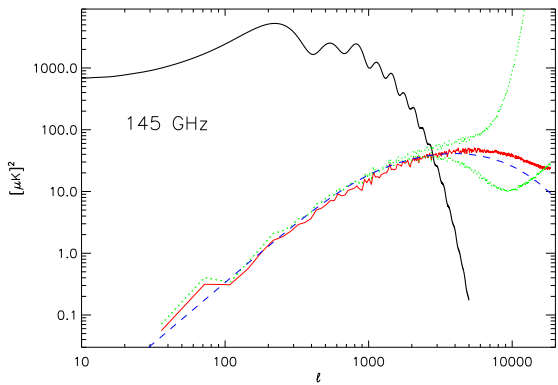}\end{minipage}
   \caption{Shown as a black solid lines is the CMB power spectrum used in the
	   simulation.  Red solid line is the total (true) SZ power spectrum
	   of  the original SZ simulation with no antenna convolution.  The
	   hybrid power spectrum constructed out of a fitting to the observed
	   ${\mathcal{N}}(S)$ is shown with dashed (blue) line.  The fiducial
	   model is used. The two (green) dotted lines show the  power spectrum
	   of the recovered SZ map ($M_{145} - M_{225}$)  (bottom green dotted
	   line) and the same  power spectrum deconvolved by the antenna (upper
	   green dotted line). The  antenna-convolved power shows an increasing
	   tail at large $\ell$'s due  to the instrumental noise. Note how the
	   hybrid power spectrum is not affected by the antenna.   }
  \label{fig_Clsimul}
   \end{flushleft}
\end{figure}

The CMB map is created from a standard WMAP-like CMB power spectrum (see
primary CMB power spectrum in figure (\ref{fig_Clsimul})).  The SZ map was
kindly provided by M. White (from his `Data archive') and  consist of an N-body
simulation normalized to the current estimates of the excess  in power of the
CMB at small scales (White 2003). The angular size of the simulation is 
$10^{\circ} \times 10^{\circ}$ with a pixel size of 35 arcsec.

After adding the CMB and the SZ maps at the three ACT frequencies we include
the beam smearing by filtering with Gaussian beams of 1.7, 1.1 and 0.93 arcmin
FWHM. In the last step we incorporate the instrumental noise assuming it is
white Gaussian with an RMS of 6, 10 and 15 $\mu$K per pixel with a pixel of 35
arcsec. The finally sky map, as would  observed by ACT, is shown in figure
(\ref{fig_ACT_fullsky}).

At this step there are a variety of methods which can be applied to recover the
SZ map from these three simulations. As an illustration we will use the most
naive approach which just takes the combination of $M_{145} - M_{225}$ as an
estimator of the SZ effect at 145 GHz. In the previous estimator, $M_{145}$ is
the ACT map from the 145 GHz channel and $M_{225}$ is the corresponding map at
225 GHz but degraded to the resolution of the 145 GHz map. Since the frequency
dependence of the CMB is constant in $\Delta T/T$ units,  the CMB disappears
completely after subtracting  $M_{225}$ to $M_{145}$.  The same happens to the
kinetic SZ effect which has no frequency dependence.  This particularly simple
estimator gives satisfactory results and has the advantage that it does not
rely on any assumption (compared to other component separation algorithms). We
should note, however,   that more elaborated methods should render better
results since, for instance,  we do not use the map at 265 GHz which could be
useful to increase the detection  rate. The final SZ map will show a small bias
toward higher (more negative) values of the SZ effect due to the fact that the
225 GHz map still contains a small positive contribution from the thermal SZ.
The recovered SZ map is a noisy estimate of the true one and we can estimate
the power spectrum of this noisy SZ map. The result is shown as the two dotted
lines in figure (\ref{fig_Clsimul}) . The bottom dotted  line is the recovered
SZ power spectrum and the top dotted line shows the beam-unconvolved power
spectrum. The true SZ power spectrum (no antenna  convolution) from the
hydro-simulation is shown as a thin solid line.  The recovered SZ power
spectrum (bottom dotted line) can not recover  all the power at small scales
($\ell > 3000$). The reason is because  our component separation algorithm
($M_{145} - M_{225}$) does not include  any beam deconvolution and the
recovered SZ map shows the damping in the small  scales due to the antenna
beam. The increase in power at $\ell > 8000$ is due  to the instrumental noise
which is not affected by the beam smearing.   One can get a better estimate of
the recovered power spectrum combining the 3 frequency  maps in an optimal way
(see e.g Mart\'\i nez-Gonz\'alez et al. 2003).

\begin{figure}
   \begin{flushleft}
   \epsfysize=6.cm
   \begin{minipage}{\epsfysize}\epsffile{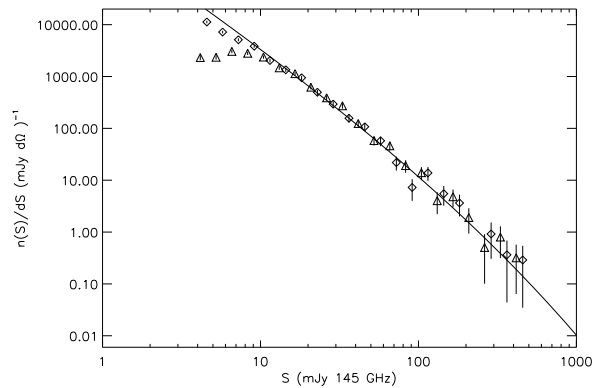}\end{minipage}
   \caption{Comparison of cluster flux counts recovered by
    \small{SEXTRACTOR}\normalsize\  (shown with triangles) with
    actual flux counts found in the SZ simulations (shown with diamonds). The
    line shows a fit to the flux counts.}
   \label{fig_true_vs_recovered}
   \end{flushleft}
\end{figure}

To construct the hybrid power spectrum we need to get an estimate of the
cluster flux  counts from the noisy recovered SZ map. We detect the clusters 
(above 3$\sigma$) by using \small{SEXTRACTOR}\normalsize\  (Bertin \& Arnouts
1996)  as our cluster detection algorithm.  In figure
(\ref{fig_true_vs_recovered}) we show the cluster number counts recovered 
after \small{SEXTRACTING}\normalsize\  the clusters from the recovered SZ map
(triangles).  As a comparison, we also show the $N(S)$ curve when we use the
same package  (\small{SEXTRACTOR}\normalsize)\  but on the original SZ
simulation (diamonds). The drop in counts in the faint end is due to the
selection function of the survey  although this selection function is ``method"
dependent. An optimized component separation plus cluster  detection algorithm
should render better results; so our results should be considered as
conservative.  Finally, once we have extracted the flux counts, we can use the 
counts (or a fit to it as shown in figure (\ref{fig_true_vs_recovered}))  to
build the hybrid power spectrum. The resultant hybrid power spectrum (for our
fiducial model A in table 1)  is shown as the dashed line in figure
(\ref{fig_Clsimul}).  One can also see that for $100 < \ell < 3000$, the hybrid
power spectrum constructed out of ${\mathcal{N}}(S)$ very well mimics the total
power spectrum from the hydro simulations. The small excess in power at  $\ell
\in (100,3000)$ may be due to an overestimation in the fit of the real
underlying  $N(S)$ curve in the bright end. 

In figure (\ref{fig_Clsimul}), notice that at $\ell > 3000$, the hybrid power
spectrum does not agree with the actual power spectrum (for the cluster model
considered here).  It is at these high $\ell$-values that the effect of gas
physics is dominant. Thus, once the low $\ell$ part (say $\ell < 2000$) is
fixed by the hybrid power spectra (which entirely depends on the observed flux
counts), one can use the observed high $\ell$ part of the observed SZ power
spectra to constrain cluster structure and evolution. That is, instead of using
the fiducial model for the cluster parameters, we chose such cluster parameters 
so as to match the hybrid power spectrum to the true power spectrum.

\section{Using the hybrid power spectrum to study cluster physics}
\label{sec_clusterphys}

\begin{figure}
   \begin{flushleft}
   \epsfysize=6.cm
   \begin{minipage}{\epsfysize}\epsffile{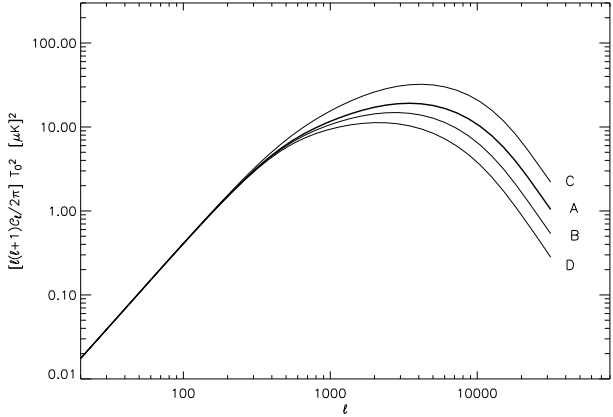}\end{minipage}
   \caption{
            Sensitivity of the hybrid power spectrum to the
            cosmological model. The models are described in Table 1 and in the text.
            The hybrid power is build from the observed curve corresponding to 
            $\sigma_8 = 0.8$ in figure (\ref{fig_Ns}). 
            The fiducial model A is shown in bold line.
           }
   \label{fig_hybridcosmo}
   \end{flushleft}
\end{figure}

In this section we come to the central point of the paper which is to  present
the hybrid power spectrum as a tool to study cluster physics. We first 
demonstrate that the hybrid power spectrum  is weakly dependent on the
cosmological model. From equation (\ref{eqn_Cl_h}), we expect the hybrid power
spectrum to be less sensitive to $\sigma_8$. This is evident  in figure
(\ref{fig_hybridcosmo}) where we have plotted the hybrid power spectrum while
changing the cosmological parameters only. The diluted influence of the
cosmological parameters on the hybrid power spectrum is  striking when one
compares the figures (\ref{fig_hybridcosmo}) \& (\ref{fig_totalcosmo}). At low
$\ell$'s (large scales) the  hybrid power spectrum is completely insensitive to
the cosmological  model. This is because at large scales, all the clusters
behave as  unresolved point sources and their power spectrum can be determined 
directly from the cluster flux counts (which is fixed by the  observations). 
The power spectrum for the different cosmological models  only differs at small
scales where the redshift distribution of the  clusters sizes (through the
angular diameter distance) and fluxes (c.f equation \ref{eqn_totalS})  plays a
part in determining the  shape of the SZ power spectrum. Also, changing the 
cosmology will change the shape of the $N(S,z)$ surface in equation
(\ref{eqn_nSz}) but since $N(S,z)$ is normalized to $\mathcal{N}(S)$, the
dependency on the cosmological  model is diluted. The end result of the
exercise is that the SZ power spectrum loses its strong  dependence on
$\sigma_8$ and $\Omega _m$. For example, comparing models A \& C, we see that
the  two power spectra differ by a factor of two or less at $\ell > 1000$,
whereas  if we had not used the extra information from flux counts we would get
the  power spectra differing by factors of four or more for a larger
$\ell$-range.  In essence, we have mitigated the influence of cosmology on the
SZ power  spectra making it suitable to study gas physics. One can still argue
that we can use the hybrid  power spectrum to constrain the cosmological model
as it in the case of the  standard power spectrum. However, in comparison to
cosmology  the hybrid power spectrum is now more sensitive to the cluster
physics  than in the case of the standard power spectrum making the hybrid
power  spectrum a valid tool for studies of cluster physics.

\begin{figure}
   \begin{flushleft}
   \epsfysize=6.cm 
   \begin{minipage}{\epsfysize}\epsffile{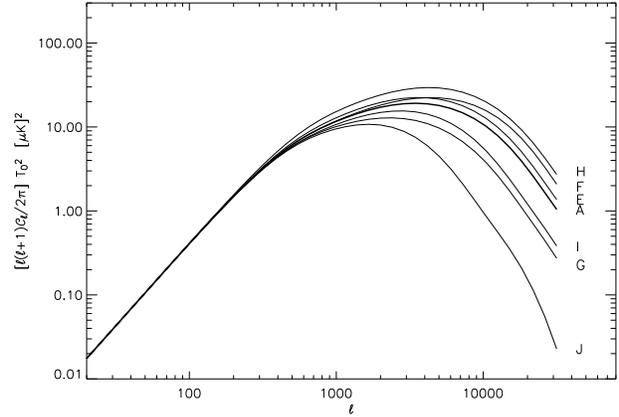}\end{minipage}
   \caption{
            Sensitivity of the hybrid power spectrum to the 
            cluster physics. The models are described in Table 1 and in the text. 
The normalization is the same as in figure (\ref{fig_hybridcosmo}). The fiducial 
model A is shown in bold line.}
   \label{fig_hybridphys}
   \end{flushleft}
\end{figure}

Next, we explore the capabilities of the hybrid power spectrum as a probe of
the  intracluster medium. We illustrate these capabilities in figure
(\ref{fig_hybridphys}) where we show how  the hybrid power spectrum behaves
once we change the cluster structural parameters  for a survey which gives a
particular ${\mathcal{N}}(S)$ (with cosmological parameters  taken from the
fiducial model). A comparison of the sensitivity of the hybrid power spectrum
to changes in cosmology  (figure \ref{fig_hybridcosmo}) and changes in the
cluster parameters (figure \ref{fig_hybridphys}) shows that the power spectrum
is  more sensitive to the cluster structure and evolution. On the contrary 
(see figures \ref{fig_totalcosmo} \& \ref{fig_totalphys}), for the standard 
total SZ power spectrum, both cosmology and cluster physics have comparable
influence  on the power spectrum. Thus at high $\ell$-values, for the standard
power spectrum it is difficult to disentangle the effects of cosmology and
cluster physics whereas for the hybrid power spectrum this degeneracy is
considerably weakened. As an example, a change of 1\% in  $\sigma_8$ changes
the standard power spectrum by $\approx 8$\%  at $\ell \approx 4000$. A similar
change (1 \%) in the parameter $\phi$ (temperature redshift evolution
parameter) changes the standard power spectrum by $0.82$ \% at $\ell \approx
4000$ whereas for the hybrid power spectrum the changes are 2.5 \% (in $\sigma
_8$) and 0.85 \% (in $\phi$)  respectively. A similar change in the two
parameters at angular resolutions probed by ALMA (say, $\ell = 30000$), 
changes the SZ power from 7.5 \% and 2 \%(standard power spectrum)  to 3.8 \%
and 1.5 \% (hybrid power spectrum) for $\sigma_8$ and $\phi$. Although these
numbers  are calculated for changes of 1 \% in $\sigma _8$ and $\phi$, we
should  note that the current uncertainty in parameters like $\phi$ or $\psi$
are  large (upto 100 \% ). Whereas, theoretically the self-similar model
predicts certain redshift dependencies (like $\phi=1$ for the M-T relation; Kaiser 1986), observationally there is still no
concurring evidence (Ettori etal 2003 and references within). Uncertainties in
the cosmological parameters are, however, much smaller (for example, WMAP gives
an uncertainty of about 5\% in $\sigma_8$ and 15 \% in $\Omega_M$; Bennett etal 2003). Thus, the
resulting hybrid power spectrum becomes a powerful probe of the cluster
physics.

\begin{figure}
   \begin{flushleft}
   \epsfysize=6.cm
   \begin{minipage}{\epsfysize}\epsffile{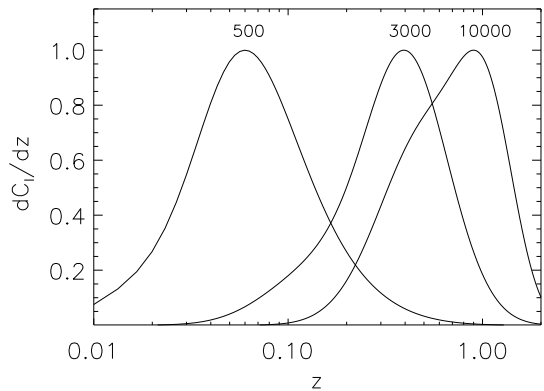}\end{minipage}
   \caption{
            Contribution to the $\ell$-multipole in the SZ power spectrum as 
            a function of the redshift of the sources 
            Three different multipoles are shown, $\ell = 500, 3000, 10000$. 
            The higher the multipole $\ell$, the more dependent is the power 
            spectrum with the high-$z$ population.
           }
   \label{fig_dCl_dz}
   \end{flushleft}
\end{figure}

It is  easy to understand the sensitivity of the hybrid power spectrum to the
underlying cluster physics.  To give an example, lets take the cluster
temperature redshift evolution  parameter $\phi$. Since the high multipoles 
are dominated by distance clusters (see figure \ref{fig_dCl_dz}),  if $\phi$ is
larger, distant clusters will appear hotter and consequently the  power
spectrum at small scales will be larger. This is why model A has more power  at
large $\ell$'s than model G. If we take model H, here the core radius is
smaller  for all clusters than in model A, consequently the clusters will look
more compact and  this will increase the power at small scales. Model J, on the
other hand, will have  larger cluster sizes than model A (specially at high
redshifts) so its power will  be smaller at small scales. 
Since evolution in cluster structure differentiates strongly the high redshift 
population w.r.t the nearby clusters, the high multipole moments of $C_\ell$ are 
very sensitive to any evolution: consequently one can strongly constrain any 
cluster structure evolution with the hybrid power spectrum.

The power of SZ power spectrum as a probe of cluster gas physics has already
been realized  (Majumdar 2001). Especially, it is seen that high $\ell$-values
SZ power spectrum is very sensitive to non-gravitational processes (Komatsu \&
Kitayama 1999, Holder \& Carlstrom 2001). This feature becomes more noticeable
when the high mass clusters are subtracted from the surveys. However, the
cosmology-gas physics degeneracy were not taken into account and  the
cosmological  parameters were fixed from other observations (say, primary CMB).
The novelty of this work is that we dilute the  cosmological dependence of the
SZ power spectrum using complementary information from the same survey itself.
Moreover, we do not have to worry about cluster redshifts.

\subsection{Sources of systematics}
The main source of
systematic error in the construction of the hybrid power spectrum  comes from
our partial knowledge of the ${\mathcal{N}}(S)$ curve. This partial  knowledge
may be due to the lack of enough statistics or a poor sensitivity  in the
instrument. A survey covering a small area in the sky will contain few
clusters  per flux bin. This will produce a noisy estimate of the
${\mathcal{N}}(S)$ curve.  A small survey area will also increase the error
bars due to cosmic variance. In  particular, the latter can have a significant
effect in the high flux interval since we  expect bright clusters to be less
common than the faint ones. A lack of bright clusters  in the sample (or
similarly an underestimated ${\mathcal{N}}(S)$ curve at high fluxes)  will lead
to an underestimated hybrid power spectrum in large scales (small $\ell$'s). 
Consequently, by subtracting the hybrid power spectrum from the total
(observed) power  spectrum one can get a {\it biased} estimate of the
correlation term if the ${\mathcal{N}}(S)$ curve is not correctly estimated  in
the high flux regime. As an immediate example, we see this error in estimation
has given us hybrid power spectrum with slightly more power than the total
power spectrum at large scales (see figure (\ref{fig_Clsimul})). One can try to
bypass this systematic by using a fit to the ${\mathcal{N}}(S)$ curve to high
fluxes. However, in order to have a good fit it is necessary to have a large
number of bright  clusters.  On the other hand, a
poor sensitivity of the instrument will reduce the number of faint  clusters
detected in the survey. The faint clusters will contribute to the hybrid power 
spectrum in the high $\ell$ regime (where the sensitivity of the hybrid power
spectrum  with the cluster physics is more relevant). Moreover, the cluster
catalog (and hereby  the ${\mathcal{N}}(S)$ curve) will be the product of a
processing of the raw data  (e.g component separation) which usually introduce
a selection function in the cluster  catalog. Understanding the selection
function is fundamental specially at the faint end of the catalog where
the selection function is expected to change dramatically. Usually, one
should fit the ${\mathcal{N}}(S)$ curve only using those  data points for which
the selection function is well understood.  Then one can extrapolate  the fit
to lower (or higher) fluxes and build $C^H_\ell$ from the fitting.

\section{Discussion and summary}
\label{sec_summary}

In the previous sections we have demonstrated that cosmology as well as cluster
gas structure  and evolution simultaneously shapes the thermal SZ power
spectrum from clusters of galaxies.  The power spectrum depends strongly on
certain cosmological parameters (like $\sigma_8$) and  on any evolution, if
present, thus leading to a cosmology-gas physics degeneracy. One can take
priors from external observations, for example fix the cosmological model from
primary CMB,  and use the power spectrum to probe gas physics. However, it is
possible to combine complimentary  information from the same survey itself to
mitigate the influence of cosmology in order to study  gas physics.

To do so, we have formulated a different way of looking at the SZ power
spectrum by constructing  the hybrid power spectrum. Given that any future
large yield SZ survey would also detect many  thousands of clusters, we would
have a flux counts of the objects detected in the survey. At the  same time,
temperature fluctuations from all the survey pixels would be used to construct
the  total SZ power spectrum. We have shown that one can rewrite the Poisson
part of the SZ power  spectrum using the information available from the SZ flux
counts. Since the Poisson power spectrum  dominates, in general, over the
clustering power spectrum, the resultant hybrid power spectrum  represents the
total SZ power spectrum sufficiently well up to $\ell \sim 2000$. We also used 
results from numerical simulations to test and verify our analytical results.
Since both the  SZ power spectrum and the flux counts represent the same
underlying cosmology, the hybrid power  spectrum is, in essence, normalized to
the background cosmology. Once the cosmology-gas physics  degeneracy is
diluted, we are left with modeling of the gas physics to understand the
observed  total SZ power spectrum at higher multipoles.

We have used a simple model of the cluster gas structure and evolution as well
analytical results  for the mass function in our analytical model. This is
sufficient for the present work which is {\it exploratory} in nature. Needless
to say, practical application of the method would need better  modeling of the
complex cluster structure using N-body hydro simulations or with more
complicated  analytical models (suitable to fit simulation results).
Cosmological simulations of large yield of  clusters mimicking upcoming surveys
are still naive in the sense that they still need some  sort of
phenomenological approach to model the gas physics (especially non-adibaticity)
and has to make simplifying assumptions  as to the cluster structure. At the
same time simulating any non-standard cluster evolution is  non-trivial and
either done too simplistically or is put in by-hand in the simulations. It is
this  very cluster structure and evolution that we are able to probe by
studying the SZ power spectrum  at high $\ell$-values.  Note, that in figure
(\ref{fig_Clsimul}), at multipoles  ($\ell > 2000$) the hybrid power spectrum
defers from the actual power spectrum. This, of  course, should be the case
since the cluster structure (c.f. equations (\ref{eqn_TM}) \& (\ref{eqn_RM})) 
used in constructing $C^H_l$ is very different from the gas physics used in the
SZ N-body simulations.  However, it must  be kept in mind that a part of the
difference may arise from increase in power  at high $\ell$'s due to presence
of substructures in the clusters which are naturally captured  in any N-body
simulations but not modeled analytically. 

The ability to construct $C^H_l$ depends crucially on our capability to get the
cluster flux counts. The main source of systematic error in the construction of
the hybrid power spectrum  comes from our partial knowledge of the
${\mathcal{N}}(S)$ curve. This partial  knowledge may be due to the lack of
enough statistics or a poor sensitivity  in the instrument. 
It is also important to understand that processing of the raw data can
introduce a selection function in the cluster catalog. 
 Usually, one should fit the ${\mathcal{N}}(S)$ curve only using those  data
points for which the selection function is well understood and then
extrapolate  the fit to lower (and/or higher) fluxes. This extrapolation will
have little effect in the constructed hybrid power spectrum if the survey is
large enough such that it contains enough bright clusters and if its
sensitivity is good enough to recover clusters with fluxes as low as a few mJy
(say, at 145 GHz). Future SZ cluster surveys (like ACT/SPT/APEX-SZ) satisfy
both these conditions such that the hybrid power spectrum can be a useful tool
in these surveys.

The main conclusion of this work can be summarized by comparing figures 
(\ref{fig_hybridcosmo} \& \ref{fig_hybridphys}) with  figures 
(\ref{fig_totalcosmo} \& \ref{fig_totalphys}). By using the hybrid power 
spectrum instead of the standard power spectrum, one can dilute the dependency 
of the power spectrum on the cosmological model and concentrate on studies  of
the cluster physics. The precision  in the determination of the cosmological
parameters has increased dramatically  in recent times. However, much
uncertainties remain in studies of  cluster physics. Better handle on cluster
structure and evolution becomes even more challenging when one takes in to
account the cosmology-gas physics degeneracy in any such studies.  The hybrid
power spectrum helps to soften this problem by diluting much of this degeneracy.

The prospect of probing gas physics using SZ power spectrum would depend on our
ability to measure  the SZ power spectrum as very high $\ell$-values. This sub
arc-min scale has the distinct advantage  of the primary CMB contribution being
negligible. However one has to be careful in eliminating other  possible
sources of secondary CMB anisotropies that may introduce further systematic
uncertainties. In multifrequency experiments, an appropriate SZ reconstruction
algorithm should be  able to recover the true SZ power spectrum up to the
resolution limit  of the experiment  

Finally, all future high sensitivity SZ observations have to
tackle the noise introduced by unresolved  point sources. The spectral
dependence of the thermal SZ effect would be helpful in eliminating many  of
such systematics. Although not studied in this work, another interesting
application of the hybrid power  spectrum is to perform consistency checks on
the excess in power in single frequency  CMB experiments.  By this we mean that
future CMB experiments will observe an excess in the power spectrum  of CMB
fluctuations at $\ell > 2000$. This excess will be due in part to galaxy
clusters,  and in part to non-removed point sources (see figure 2 in White \&
Majumdar 2003). Now, at $\ell \sim 2000$ the hybrid power spectrum is a decent
representation of the total SZ power spectrum without worrying much about
cluster physics. If there is an estimation of the $\mathcal{N}(S)$ curve for
the survey, then one can estimate SZ power due to clusters at $\ell \sim
2000$.  Any excess power can be interpreted as due to unresolved
point sources. 

In spite the stringent requirements needed to make SZ observations from the
upcoming surveys probe of  cluster physics, the rewards reaped would
undoubtedly be great. In addition to new high  resolution targeted
observations, the statistical study of the SZ sky from cosmological
distribution  of clusters may provide the next leap in the study of the
structure and evolution of galaxy clusters.

\section{Acknowledgments} 
The authors would like to thank Martin White for
providing the SZ maps and the group catalogs. We also would like to thank Max
Tegmark, Ue-Li Pen,  Gil Holder, Martin White and Maija Jespersen for
discussions and suggestions.  The work of JMD was supported by the David and
Lucile Packard Foundation and the Cottrell Foundation.

\newpage
{\bf {\LARGE \noindent Appendix A}}\\
{\bf {\Large \noindent The hybrid power spectrum for X-ray surveys}}\\

\normalsize
The computation of the hybrid power spectrum for the case of X-ray surveys 
can be done following the same procedure as in the SZ case. 
We start with a measured cluster number counts,  ${\mathcal{N}}(S)$ 
(where now the fluxes are given in ergs s$^{-1}$ cm$^{-2}$). 
The fluxes can be connected with the mass through the equations;
\begin{equation}
S = \frac{L_x}{4 \pi D_l^2}
\label{equation_Sx}
\end{equation}
\begin{equation}
L_x = L_o M^{\alpha} (1+z)^{\phi}
\label{equation_Lx}
\end{equation}
where $L_x$ is the cluster X-ray luminosity and $D_l$ is the luminosity 
distance. The constant $L_o$ will contain all the units and the 
conversion factors (band correction, conversion between cts/s and 
ergs/s).
Equation (\ref{equation_Lx}) will play the same role as the 
$T-M$ relation in the case of a SZ survey. 
The final ingredient we need to compute the X-ray hybrid power spectrum 
is the function $f(\ell,M,z)$ which will differ to the one 
described in section \ref{sec_modelling}. 
For the case of X-ray images, the 2D profile of a cluster will look 
different since is proportional to the integral along the line of sight 
of the density squared. 
The fourier transforms of these 2D profiles can be fitted rendering;
\begin{equation}
f(\ell,M,z) = \frac{1}{2}\left(\exp(-\xi_{l,r_c}) + \exp(-\sqrt{\xi_{l,r_c}}) \right)
\label{eqn_fit_Cl}
\end{equation}
with, 
\begin{equation}
\xi_{l,r_c} = l^2 r_c^{1.5/(0.815 + 0.35r_c)}
\end{equation}
where the core radius, $r_c$, is given in rads (see Diego et al. 2003).

\end{document}